\documentclass[aps,preprint,showpacs,showkeys]{revtex4}%
\usepackage{amsmath,bm}
\usepackage{amsmath}
\usepackage{amsfonts}
\usepackage{amssymb}
\usepackage{graphicx}%
\input{psfig.sty}

\begin{document}
\date{\today }
\title{High Spins Beyond Rarita-Schwinger Framework}
\author{Mariana Kirchbach$^{1}$, Mauro Napsuciale$^{2}$}
\affiliation{$^1$Instituto de F\'{\i}sica, \\
         Universidad Aut\'onoma de San Luis Potos\'{\i},\\
         Av. Manuel Nava 6, San Luis Potos\'{\i}, S.L.P. 78290, M\'exico}
\affiliation{$^{2}$Insituto de F\'{\i}sica, Univ.\ de Guanajuato, 
Lomas del Bosque 103,
Fracc.\ Lomas del Campestre, 37150 Leon, Guanajuato, M\'exico}

\begin{abstract}
We study the eigenvalue problem of the squared Pauli-Lubanski vector,
${\mathcal W}^{2}$, in the
Rarita-Schwinger representation space and derive from it that the $\left(
-s(s+1)m^{2}\right)  $ subspace with $s=3/2$, i.e. 
spin 3/2 in the rest frame, is pinned
down by the one sole Klein-Gordon like equation, 
$\left[  (p^{2}-m^{2})g_{\alpha\beta}-\frac
{2}{3}p_{\beta }p_{\alpha }- \frac{1}{3}(p_{\alpha}\gamma_{\beta}+p_{\beta
}\gamma_{\alpha})\not p  +\frac{1}{3}\gamma_{\alpha}\not p  \gamma_{\beta
}\not p  \right]  \psi^{\beta}=0$. Upon gauging this 
${\mathcal W}^{2}$ invariant subspace of $\psi_{\mu}$ 
is shown to couple to the electromagnetic 
field in a fully covariant fashion already at zeroth order
of $1/m$ and with the correct gyromagnetic factor of 
$g_{\frac{3}{2}}=\frac{2}{3}$. 
The gauged  equation is hyperbolic and hence free from the 
Velo-Zwanziger problem of 
acausal propagation within an electromagnetic field
at least to that order.

\end{abstract}
\maketitle

\section{Introduction}
High spin particles occupy an important place in
theoretical physics. For the first time they were observed as
resonant excitations in pion-nucleon scattering. The Particle Data
Group \cite{PART} lists more than thirty non-strange baryon 
resonances with spins ranging from 3/2 to 15/2, and more than 
twenty strange ones with spins from 3/2 to 9/2. 
Baryon resonances have been extensively investigated in the past
at the former Los Alamos Meson Physics Facility (LAMPF), 
and at present their study continues at 
the Thomas Jefferson National Accelerator Facility (TJNAF)
\cite{Burk_Lee}.
Such particles are of high relevance in the description
of photo- and electro-pion production off proton, where they
appear as intermediate states, studies to which the Mainz Microtron (MAMI) 
devotes itself since many years \cite{MAMI}.
Search for high spin solutions to the QCD lagrangian
has been recently reported by the Lattice collaboration in
Ref.~\cite{32llatt}. Moreover, also the twistor formalism has been 
employed in the construction of high spin fields \cite{32_twistor}.
Integer high spins meson resonances with spins ranging from 0 to 6 
can have importance in various processes revealing the fundamental 
features of QED at high energies  such like pair production \cite{Kruglov}.  
However, the most attractive high spin fields 
appear in proposals for physics beyond
the standard model which invoke supersymmetry \cite{Kaku}
and contain gauge  fields of fractional spins such as the
gravitino- the supersymmetric partner of the ordinary spin-2
graviton. Supersymmetric theories open the
venue to the production of fundamental spin $\frac{3}{2}$ particles 
at early stages of the universe, whose understanding can play an 
important role in its evolution \cite{Verdi}.
 
The description of high spins takes its origin from Refs.~\cite{FiePau},
\cite{RS}, \cite{Weinbrg65} who suggest to consider any fractional 
spin-$s$ as the highest spin
in the totally symmetric rank-$(s-1/2)$ Lorentz tensor with Dirac spinor 
components, $\psi_{\mu_{1}...\mu_{s-\frac{1}{2}}}$. For spin 3/2 one has to
consider the four-vector--spinor, $\psi_{\mu}$,
\begin{eqnarray}
\psi_{\mu}&=&A_{\mu}\otimes \psi\simeq 
\left( \frac{1}{2},\frac{1}{2}\right)\otimes
\left[ \left( \frac{1}{2},0\right)\oplus 
\left( \frac{1}{2},0\right)\right]\, ,
\label{drct_prdct}
\end{eqnarray}
the direct product between the four vector, $A_{\mu}$, 
and the Dirac spinor, $\psi $.
 As shown in \cite{MC}, in the latter case it is possible to derive 
the free particle
equations from a family of lagrangians depending on a free parameter $A$,
\begin{equation}
\mathcal{L}(A)=\overline{\psi}^{\mu}\left[p_{\alpha}\Gamma_{\mu\quad\nu}%
^{\quad\alpha}(A)-m g_{\mu\nu}\right]\psi^{\nu}\,,
\label{RS_Lgr}%
\end{equation}
where%
\begin{align}
p_{\alpha}\Gamma_{\mu\quad\nu}^{\quad\alpha}(A)\psi^{\nu}  &  =\not p
\psi_{\mu}+B(A)\gamma_{\mu}\not p  \,\gamma\cdot\psi+A(\gamma_{\mu}p\cdot
\psi+p_{\mu}\gamma\cdot\psi)
+C(A)m\gamma_{\mu}\gamma\cdot \psi\, 
\,,\label{Teil_1}\\
& \label{Teil_2}\\
A\neq\frac{1}{2},  &  \,B(A)\equiv\frac{3}{2}A^{2}+A+\frac{1}{2},\quad
C(A)=3A^{2}+3A+1. \label{Teil_3}%
\end{align}
The wave equation following from this lagrangian is obtained as
\begin{equation}
(\not p  -m)\psi_{\mu}+A\,(\gamma_{\mu}p\cdot\psi+p_{\mu}\,\gamma\cdot
\psi)+B(A)\,\left(  \,\gamma_{\mu}\not p  \,\gamma\cdot\psi\,\right)
+C(A)\,m\gamma_{\mu}\gamma\cdot\psi=0\,, \label{L(A)}%
\end{equation}
which for $A=-1$ can be written in a compact form as
\begin{equation}
(i\varepsilon_{\mu\nu\beta\alpha}\gamma^{5}\gamma^{\beta}p^{\alpha}%
-mg_{\mu \nu} +m\gamma_{\mu}\gamma_{\nu})\psi^{\nu}=0. \label{L(A=-1)}%
\end{equation}
Equations of this type  are equivalent to 
\begin{eqnarray}
\frac{1}{2m}(\not p  + m)\psi_{\mu}  &  =& \psi_{\mu} \,,
\label{RS_Dirac}\\
(-g_{\mu\nu} + \frac{1}{m^{2}} p_{\mu } p_{\nu})\psi^{\nu}
&  =& -\psi_{\mu}\,,\label{RS_Proca}\\
\gamma^{\mu}\psi_{\mu}  &  =&0\,, \label{RS_What}%
\end{eqnarray}
known as the Rarita-Schwinger \cite{RS} framework.
Notice that for the sake of convenience of the point we are going to 
make in the next section,
we here wrote the respective Dirac and Proca equations
(\ref{RS_Dirac}), and (\ref{RS_Proca}) in terms of covariant
projectors picking up spin-1/2$^{+}$ and spin-1$^{-}$ states, 
respectively. Spin 3/2$^{+}$ needs an axial four vector.
The freedom represented by $A$ reflects invariance under the so called
point transformations which
mix the two spin 1/2 sectors residing in the RS representation space besides
spin 3/2. Unfortunately for interacting fields this freedom gives rise to 
undetermined "off-shell" parameters yielding an ambiguous 
theory \cite{NL,Benm}.
As we will see below, this is not to remain the only disadvantage 
of the RS framework.
Over the years, Eqs.~(\ref{RS_Dirac})-(\ref{RS_What}) 
have been widely  applied in hadron physics to the description
of mainly the $\Delta (1232)$ and occasionally the $D_{13}(1520)$ 
resonances and their  contributions to various processes.
A recent account of effective theories for the calculation of properties  of 
light hadrons using the Rarita-Schwinger formalism  along the line of Chiral 
Perturbation Theory ~\cite{32_CHPT} can be found in \cite{Hemmert}. 

Yet, a detailed study  of Eqs.~(\ref{RS_Dirac}), (\ref{RS_Proca}), 
and (\ref{RS_What}) reveals that the 
Rarita-Schwinger framework suffers  two
more fundamental weaknesses. These are:
\begin{enumerate} 
\item The interacting \underline{quantum} spin-3/2 field  is 
plagued by several problems ranging from
loss of constraints to a failure to propagate, 
observations reported by
Johnson and Sudarshan in Ref.~\cite{sudarshan1},
\item The wave fronts of the \underline{classical} solutions of
the Rarita-Schwinger spin-3/2 equations 
suffer acausal propagation within the electromagnetic environment,
an observation due to  Velo and Zwanziger \cite{VZ1}, \cite{VZ2}.
\end{enumerate}
Although several remedies to Eq.~(\ref{RS_Lgr}) 
have been suggested over the years
(Ref.~\cite{Pasc} being the most recent),
none of them resulted in a covariant, unique,  parameter-- and
pathology free high spin wave equation.
The reason for that, as we see it, has been to not have 
worried sufficiently about the 
fundamental principles behind the high-spin description.

It is the  goal of the present study to
\begin{enumerate}
\item unveil importance of Poincar\'e invariance
and its Casimir operators,
the squared four momentum, $P^{2}$, and the squared Pauli-Lubanski
vector, ${\mathcal W}^{2}$, for the dynamics and unique identification of
the spin-degrees of freedom in $\psi_{\mu_{1}...\mu_{s-\frac{1}{2}}}$
with the special emphasis on spin-3/2 in $\psi_{\mu}$,
\item derive a unique, parameter free, wave equation for spin-3/2
 that preserves the quality of the Rarita-Schwinger field
of having correct electromagnetic couplings such as
a correct gyromagnetic factor and which is furthermore free from the 
Velo-Zwanziger problem at least to leading (zeroth) order 
with respect to $\frac{1}{m}$.
\end{enumerate}

The literature devoted to the description of high spins by
means of the Rarita-Schwinger formalism is overwhelming and contains creative
treatments at many occasions. We here focus on suggesting a solution
to the Velo-Zwanziger problem and aim to keep the presentation as concise
as possible. 
In view of this goal, we restrict ourselves to quote only few works 
which are indispensable for the point we have to make, hence we leave aside 
a vast amount of otherwise interesting and important articles.

\noindent The paper is organized as follows. In the next Section we reveal the
relationship between the Rarita-Schwinger lagrangian and the eigenvalue
problem of the squared Pauli-Lubanski vector in the four-vector--spinor. In
Section III we briefly review the Velo-Zwanziger problem for completeness of
the presentation. In Section IV we present the dynamics of a spin-3/2 particle 
in terms of the Casimir operators of the Poincar\'e group, analyze the minimal 
coupling to an external electromagnetic field and the Velo-Zwanziger problem. 
The paper closes with a brief summary and has one Appendix.

\section{The $\mathcal{W}^{2}$ background to
 the Rarita--Schwinger lagrangian.}
We envisage our goal in taking a radically
new look on the old problem. We remark that although the
Rarita-Schwinger formalism is supposed to describe spin 3/2,  
unlike the Dirac (\ref{RS_Dirac}) and Proca (\ref{RS_Proca}) 
cases, none of the equations is
a manifest covariant projector onto  
the subspace in $\psi_{\mu}$ of the desired spin and parity.

The Rarita-Schwinger  formalism does not do anything more
but successively  track first the Dirac spin 1/2 piece, 
then the Proca spin 1 piece, and 
finally their {\it supposed\/} coupling to spin 3/2 
by means of the respective Eqs.~(\ref{RS_Dirac}), Eq.~(\ref{RS_Proca}), 
and Eq.~(\ref{RS_What}). In other words,
the search for spin 3/2 in $\psi_{\mu}$ is realized
along the line of ordinary  $SU(2)$ spin--angular-momentum coupling. 

\noindent
In the present work we aim to proceed differently and
identify spin 3/2  directly and in a covariant fashion.
To do so we first retreat from the $SU(2)$ labeling
and switch to Poincar\'e group labels.
Recall that according to their definition as
irreducible representations of the Poincar\'e group particles
have to be labeled by numbers that relate to
the eigenvalues of the two Casimir
invariants of the Poincar\'e group, the squared four-momentum
$P^{2}$, and the squared Pauli-Lubanski vector, ${\mathcal W}^{2}$,
 according to \cite{ED},
\begin{eqnarray}
P^{2} \Psi^{(m,s)} &=& m^{2}
\Psi^{(m,s)}\, ,\nonumber\\
{\mathcal W}^{2}
\Psi^{(m,s)}& =& 
-m^{2}s(s+1)
\Psi^{(m,s)}\, .
\label{PGR_LBL}
\end{eqnarray}
Here $m$ stands for the mass, while $\Psi^{(m,s)}$ denotes a generic
Poincar\'e group 
representation of mass $m$ and rest-frame spin $s$.
The reason for which the Pauli-Lubanski (PL) vector plays a pivotal role in 
particle classification is that the eigenvalue problem of its square 
encodes in any inertial frame the rest-frame $SU(2)$ spin (see Appendix).
We will apply Eqs.~(\ref{PGR_LBL}) to the vector-spinor and search for 
the $s=3/2$ value. The Pauli-Lubanski vector in $\psi_{\mu}$ is defined as
\begin{equation}
{\mathcal W}_{\mu}= W_{\mu}\otimes 1_{4}+1_{4}\otimes w_{\mu}\, ,
\label{PL_3_2}
\end{equation}
where $W_{\mu}$ and $w_{\mu}$ in turn stand for the Pauli-Lubanski vectors
in the four-vector and Dirac spinor spaces, respectively, while
$1_{4}$ denotes the $4\times 4$ unit matrix.
Applied to the spin $3/2$ sector of the Rarita-Schwinger
field, $\psi^{(m,3/2)}_{\alpha}$, Poincar\'{e}
invariance requires it to satisfy \cite{MK03}
\begin{eqnarray}
\left[{\mathcal W}^{2}+m^{2}\frac{15}{4}1_{4}\otimes1_{4}\right]  _
{\nu}^{\quad\eta
}\ \psi^{(m,3/2)}_{\eta}&=&0,\nonumber\\
\psi^{(m,3/2)}_{\eta}&=&\left[A\otimes \psi \right]_{\eta}^{(m,3/2)}\, .
\label{master}%
\end{eqnarray}
The action of the Casimir invariant
$\mathcal{W}^{2}$ onto the RS field now results in
\begin{equation}
\left[  \mathcal{W}^{2}\right]_{\alpha}^{\quad\beta}\psi^{(m,3/2)}_{\beta}
=\left[ w^{2} +W^{2}\right]\psi^{(m,3/2)}_{\alpha} +
2\left(  W^{\mu}\right)  _{\alpha}^{\quad\beta}
\left[ A\otimes w_{\mu}\psi \right]_{\beta }^{(m,3/2)}\,\,.
\label{Gl1}
\end{equation}
In using  Eqs.~(\ref{W2_hh}) from the Appendix
we find
\begin{eqnarray}
\left[  
\mathcal{W}^{2}\right]  _{\alpha}^{\quad\beta}
\psi^{(m,3/2)}_{\beta}
&=&\left[ 
-\frac{3}{4} 
p^{2} -2(p^{2}g_{\alpha}^{\beta}-
p^{\beta}p_{\alpha} )
\right]
\psi^{(m,3/2)}_{\beta}+
2\left(  
W^{\mu}\right)  _{\alpha}^{\quad\beta}
\left[ A_{\beta}
\otimes w_{\mu}\psi \,\right]^{(m,3/2)}
\nonumber\\
&=& -m^{2}\frac{15}{4}\psi^{(m,3/2)}_{\alpha}\, .
\label{Gl2}
\end{eqnarray}
In making use of the mass shell condition $p^{2}=m^{2}$
in Eq.~(\ref{Gl2}) 
allows to reduce it to
\begin{equation}
2\left(  W^{\mu}\right)  _{\alpha}^{\quad\beta}
\left[ A\otimes w_{\mu}
\,\psi\right]_{\beta }^{(m,3/2)} +\,m^{2}\psi^{(m,3/2)}_{\alpha}\,
=-2p_{\alpha}p\cdot\psi^{(m,3/2)}\, .
\label{ancestor}%
\end{equation}
Notice that spin 3/2 or spin 1/2 are identified by
the eigenvalues of the first term on the left hand side of 
Eq.~(\ref{ancestor}),  $-m^{2}$ versus $2m^{2}$.
In order to construct  explicitly this term which will be referred to
as the Dirac--Proca ``intertwining term''  we exploit, 
$\ \gamma^{\nu}\gamma^{\rho}=g^{\nu\rho}-i\sigma
^{\nu\rho}$, to cast $w_{\mu}$ from Eq.~(\ref{PL_Dir}) into the more
convenient form
\begin{equation}
w_{\mu}=\frac{1}{2}\gamma_{5}(p_{\mu}-\gamma_{\mu}\not p )\,.
\label{Dir_PL_final}%
\end{equation}
Substitution of Eq.~(\ref{Dir_PL_final}) into the intertwining 
term  yields%
\begin{align}
2\left( W^{\mu}\right)  _{\alpha}^{\quad\beta}   w_{\mu} &  =i\epsilon^{\mu
}\,_{\alpha}\,^{\beta}\,_{\sigma}p^{\sigma}\gamma_{5}(p_{\mu}-\gamma_{\mu
}\not p )\nonumber\\
&  =-i\epsilon^{\mu}\,_{\alpha}\,^{\beta}\,_{\sigma}p^{\sigma}\gamma_{5}%
\gamma_{\mu}\not p \,.\, \label{zw_stf_1}%
\end{align}
Substitution of Eq.~(\ref{zw_stf_1}) in Eq.~(\ref{ancestor})
results in the following equation:
\begin{equation}
(i\epsilon_{\alpha\beta\mu\sigma}\gamma^{5}\gamma^{\mu}p^{\sigma}\not p
 -m^{2}g_{\alpha\beta} +2p_{\beta }p_{\alpha})\psi^{(m,3/2)}\, ^{\beta}=0. 
\label{intf_cuad}%
\end{equation}
The resemblance between Eqs.~(\ref{intf_cuad})  and ~(\ref{L(A=-1)}) 
is hardly to be overlooked. The kinetic and mass terms are identical
(modulo the Dirac condition $p\!\!\!/ \psi_{\mu}=m\psi_{\mu}$),
while the remaining terms only represent different auxiliary conditions,
a difference that should be irrelevant at that stage as both
wave equations provide equivalent free particle descriptions.  
Through above observation the Rarita-Schwinger lagrangian unexpectedly 
acquires the status of a substitute to the complete ${\mathcal W}^{2}$ 
eigenvalue problem. 
Within the context of this new reading of the Rarita-Schwinger lagrangian
the question arises as to what extent the problems of the
traditional  high-spin description may be 
resolved in recovering the complete ${\mathcal W}^{2}$ action on 
$\psi_{\mu}$ given in Eq.~(\ref{Gl2}).

Below we will make the case that
\begin{quote}
the complete $\mathcal{W}^{2}$ eigenvalue problem gives rise to a simple,
transparent and easy to handle equation for the spin-3/2 degrees of freedom in
$\psi_{\mu}$ that has the appealing advantage to be free from the chronic
deficit of the Rarita-Schwinger framework, the acausal propagation 
in the presence of an electromagnetic field, at least to leading
(zeroth)  order in  $\frac{1}{m}$.
\end{quote}

Before presenting the mentioned equation in Section IV below, we briefly review
the Velo-Zwanziger problem in the next Section.

\section{The Velo-Zwanziger problem of linear high-spin lagrangians}

The work of Giorgio Velo and Daniel Zwanziger on the defects of the linear RS
lagrangian for interacting particles with high spins \cite{VZ1} occupies the
central place in the present study. {}For the sake of self sufficiency of the
present paper we here highlight it in brief.
Also from here onwards we will suppress upper labels $\, ^{(m,s)}$
for the sake of simplicity of the notations.

The main point of Ref.~\cite{VZ1} is that Eq.~(\ref{L(A=-1)}) provided by the
lagrangian device (\ref{Teil_1}) is not a genuine equation of motion because 
it is of 
first order but the time derivative of $\psi_{0}$ never occurs. This defect
reveals itself in multiple ways through

\begin{enumerate}
\item the complete cancellation of all $\partial_{0}\psi_{0}$ terms in
Eq.~(\ref{L(A=-1)}) for any $\mu$,

\item the complete cancellation of $\partial_{0}\psi_{\alpha}$ terms for
$\mu=0$, in which case one finds instead of a wave equation the constraint
\begin{equation}
\lbrack\vec{p}+(\vec{p}\cdot\,\vec{\gamma}-\gamma^{0}m)\vec{\gamma}]\cdot
\vec{\psi}=0\,, \label{prime_CS}%
\end{equation}

\item the absence of $\psi_{0}$ in Eq.~(\ref{prime_CS}) that leaves the
time-component of the Rarita-Schwinger field undetermined,
\end{enumerate}
In fact above deficits are due to the constraints incorporated in the
wave equations and could be tolerated if remediable upon gauging.
Velo and Zwanziger elaborate in Ref.~\cite{VZ1} the gauging procedure in
replacing $p_{\mu}$ by $\pi_{\mu}=p_{\mu}+eA_{\mu}$, and succeed in shaping
Eq.~(\ref{L(A=-1)}) to a genuine equation \footnote{Velo and Zwanziger
consider the case of $A=-1$, the $A=-\frac{1}{3}$ has been worked out in
Refs.~\cite{MC}, \cite{Weda}.}. The remedy starts with first contracting the
gauged equation successively by $\gamma^{\mu}$ and $\pi^{\mu}$ and generating
the covariant gauged constraints as 
\footnote{Notice that we are using different 
conventions for $\gamma_5$ and the Levi-Civita tensor.}
\begin{align}
\gamma\cdot\psi &  =\frac{2}{3}\frac{e}{m^{2}}\gamma^{5}\gamma
\cdot\widetilde{F}\cdot\psi\,,\label{1st_GSS}\\
\pi\cdot\psi &  =(\gamma\cdot\pi+\frac{3}{2}m)\frac{2}{3}\frac{e}{m^{2}%
}\gamma^{5}\gamma\cdot{\widetilde{F}}\cdot\psi\,, \label{2nd_GSS}%
\end{align}
and ends with substituting Eqs.~(\ref{1st_GSS},\ref{2nd_GSS})  
back into the gauged  Eq.~(\ref{L(A=-1)}). The
resulting new wave equation,
\begin{equation}
(\not \pi -m)\psi_{\mu}-(\pi_{\mu}+\frac{1}{2}\gamma_{\mu})\frac{2}{3}%
\frac{e}{m^{2}}\gamma^{5}\gamma\cdot{\widetilde{F}}\cdot\psi\,=0\,,
\label{VZ_gauged}%
\end{equation}
is a true one because it specifies the time derivatives of $\psi_{\mu}$ for
any given $\mu$. Also $\psi_{0}$ is now defined by means of Eq.~(\ref{1st_GSS}%
). The ultimate step is testing hyperbolicity and causality of
the solutions to Eq.~(\ref{VZ_gauged}) by
means of the Courant-Hilbert criterion \footnote{The Courant-Hilbert criterion
for hyperbolicity requires the determinant of the coefficient matrix obtained
by replacing the highest derivatives by $n_{\mu}$, the normals to the
characteristic surfaces, to vanish only for real $n_{0}$.}. The result is that
in general the equation is not hyperbolic and the wave front velocity of
its solutions can exceed the speed of light. Strong fields destroy
the hyperbolic character of Eq.~(\ref{VZ_gauged}).
It was found that solely sufficiently weak fields
guarantee hyperbolicity of Eq.~(\ref{VZ_gauged}). \newline
To the best of our knowledge no solution to
the Velo-Zwanziger problem has been suggested so far.
It solely has been pointed out by Hurley \cite{Hurley_D4} that 
the generalized Feynman--Gell-Mann equations
for $(s,0)\oplus (0,s)$ representations, 
\begin{equation}
(\pi ^{2}-m^{2})
\Psi^{(s,0)\oplus (0,s)}=
\frac{e}{2s} S^{\mu\nu}F_{\mu\nu}\Psi^{(s,0)\oplus (0,s)}\, ,
\label{Hurley}
\end{equation} 
are obviously manifestly hyperbolic. 
However, $(s,0)\oplus (0,s)$ states are difficult
to couple to the pion-nucleon or photon-nucleon system
due to dimensionality mismatch, a reason that represents a
serious obstacle to their application in phenomenology.

In the next Section we show that Poincar\'{e} invariance of space time
suggests a different equation in which (i) the wave fronts propagate causally,
(ii) the coupling to the electromagnetic field
appears in a fully covariant fashion already  at leading order 
with respect to $1/m$ and wears the
correct gyromagnetic factor of $g_{s}=1/s$.

\section{Beyond Rarita-Schwinger framework--consistent description of high
spins.}

The observed kinship between the lagrangian equation (\ref{L(A=-1)}) 
and the
$\mathcal{W}^{2}$ eigenvalue problem discussed in Section II above
is suggestive of the idea to test
predictive power of Eq.~(\ref{Gl2}) for the description of high spins.

\subsection{The free case}

In taking this venue we substitute Eq.~(\ref{zw_stf_1}) in
Eqs.~(\ref{Gl2}) to obtain the full $\mathcal{W}^{2}$ operator as
\begin{align}
(\mathcal{W}^{2})_{\alpha\beta}  &  =-\frac{3}{4}p^{2}g_{\alpha\beta}%
-2(p^{2}g_{\alpha\beta}-p_{\beta }p_{\alpha })-p^{2}g_{\alpha\beta}+(p_{\alpha
}\gamma_{\beta}+p_{\beta}\gamma_{\alpha}-\gamma_{\alpha}\not p \gamma_{\beta
})\not p \nonumber\\
&  =-\frac{15}{4}p^{2}g_{\alpha\beta}+2p_{\beta }p_{\alpha }+(p_{\alpha}%
\gamma_{\beta}+p_{\beta}\gamma_{\alpha}-\gamma_{\alpha}\not p \gamma_{\beta
})\not p \,.
\end{align}
Hence the associated spin 3/2 equation reads
\begin{equation}
\left[  \frac{15}{4}(p^{2}-m^{2})g_{\alpha\beta}-2p_{\beta }p_{\alpha
}-(p_{\alpha}\gamma_{\beta}+p_{\beta}\gamma_{\alpha}-\gamma_{\alpha}%
\not p \gamma_{\beta})\not p \right]  \psi^{\beta}=0. \label{w2}%
\end{equation}
If this equation is to pin down correctly the eight spin-3/2 degrees of
freedom in $\psi_{\mu}$ it has to incorporate the supplementary condition
which should be found by contracting successively with $\gamma^{\alpha}$, and
$p^{\alpha}$. In so doing we find
\begin{align}
\left[  p^{2}-5m^{2}\right]  \gamma\cdot\psi &  =0,\label{mist1}\\
\left[  p^{2}-5m^{2}\right]  \,\,p\cdot\psi &  =0. \label{mist2}%
\end{align}
This result reflects certain occasional confusion in the $s$ value. 
In the concrete case
we are facing the  twofold ambiguity in the decomposition of the
$\mathcal{W}^{2}$ eigenvalues according to
\begin{equation}
-\frac{15}{4}m^{2}={\Big\lbrace}%
\begin{array}
[c]{c}%
-\frac{3}{2}(\frac{3}{2}+1)m^{2}\\
-\frac{1}{2}(\frac{1}{2}+1)5m^{2}%
\end{array}
, \label{uhuh}%
\end{equation}
that attributes a mass of $\sqrt{5}m$ to the spin-1/2 fields $\gamma\cdot\psi$
and $p\cdot\psi$, respectively \cite{Selim}. In other words, 
at times the $\mathcal{W}^{2}$ eigenvalue by itself ceases to 
fix unambiguously the spin unless the eigenvalue of 
the first Casimir invariant, $p^{2}$, has been specified.
The problem is resolved upon setting the Dirac sector on mass shell, i.e. in
assuming
\begin{equation}
w^{2}=-\frac{3}{4}p^{2}=-\frac{3}{4}m^{2}. \label{Dir_OMS}%
\end{equation}
In substituting the latter equation in Eq.~(\ref{w2}) amounts to
the following free Klein-Gordon like spin 3/2 equation
\begin{equation}
 (W^{2})_{\alpha\beta}\psi^{\beta }+3m^{2}\psi_{\alpha}
+2(W^{\mu})_{\alpha\beta} w_{\mu}\psi^{\beta}=0\,,
\label{pre_remd2} 
\end{equation}
the explicit form of which reads
\begin{eqnarray}
\left[  (p^{2}-m^{2})g_{\alpha\beta}-\frac{2}{3}
p_{\beta}p_{\alpha }-\frac{1}{3}(p_{\alpha}%
\gamma_{\beta}+p_{\beta}\gamma_{\alpha}-\gamma_{\alpha}\not p \gamma_{\beta
})\not p \right]  \psi^{\beta}&=&0. \label{platino}%
\end{eqnarray}
In now contracting this equation successively by $p^{\alpha}$ and
$\gamma^{\alpha}$ one recovers the standard auxiliary conditions
\begin{align}
p\cdot\psi &  =0\,,\label{remd1}\\
\gamma\cdot\psi &  =0. \label{remd2}%
\end{align}
Introducing these constraints into Eq.~(\ref{platino}) amounts to
the Klein-Gordon condition for all the field $\psi_{\mu}$ components
\begin{equation}
(p^{2}-m^{2})\psi_{\mu}=0,
\end{equation}
as it should be.
Our free Eq.~(\ref{platino}) has one problem in common with the
RS framework--  it is also  not a genuine equation 
because the highest time derivative of $\psi_{0}$ never occurs.
On the other side, it can be shown that 
its $\mu =0$ component does not amount to a constraint
as in Eq.~(\ref{prime_CS}) but fixes $\psi_{0}$ and 
also contains first order time derivatives.
In the next subsection we will couple  Eq.~(\ref{platino}) 
to electromagnetism.

\subsection{The interacting case}

In now replacing $p^{\mu}$ by $\pi^{\mu}=p^{\mu}-eA^{\mu}$ 
in Eq.~(\ref{pre_remd2}), the gauged equation is rewritten as
\begin{equation}
\left[  (\widehat{W}^{2})_{\alpha\beta} +3m^{2}g_{\alpha\beta}+
2(\widehat{W}^{\mu
})_{\alpha\beta}\widehat{w}_{\mu}\right]  \psi^{\beta}=0\,,
\label{gauged_hidden}
\end{equation}
where hat-quantities are obtained from the bare ones by the
replacement $p^{\mu}$ $\longrightarrow$ $\pi^{\mu}$. 
The gauged Dirac-Proca intertwining term is now calculated as
\begin{eqnarray}
2(\widehat{W}^{\mu})_{\alpha\beta}\widehat{w}_{\mu}=(i\epsilon_{\quad
\alpha\beta\sigma}^{\mu}\pi^{\sigma})(\frac{i}{2}\gamma_{5}\sigma_{\mu\rho}%
\pi^{\rho})&=&-\widehat{T}_{\alpha\beta}\not \pi -e\gamma_{5}\widetilde
{F}_{\alpha\beta}\,,
\label{DP_ITW_G}\\
\widehat{T}_{\alpha\beta}&=&i\epsilon_{\alpha\beta\mu\sigma}\gamma
_{5}\gamma^{\mu}\pi^{\sigma}\, ,
\label{T_ab}
\end{eqnarray}
where $\widetilde{F}_{\alpha\beta}$ is the dual
to the electromagnetic field strength tensor,
$\widetilde{F}_{\alpha\beta}=\frac{1}{2}\epsilon
_{\alpha\beta\mu\sigma}F^{\mu\sigma}$, while $e$ stands for the electric
charge of the field. Finally, $[\pi_{\alpha},\pi_{\beta}]=ieF_{\alpha\beta}$,
with $F_{\alpha\beta}=\partial_{\alpha}A_{\beta}-\partial_{\beta}A_{\alpha}$.
Putting everything together we obtain the following equation:%
\begin{equation}
\left[  \widehat{T}_{\alpha\beta}\not \pi +(2\pi^{2}-3m^{2})g_{\alpha\beta
}-2\pi_{\alpha}\pi_{\beta}+e\gamma_{5}\widetilde{F}_{\alpha\beta}%
+2ieF_{\alpha\beta}\right]  \psi^{\beta}=0\,, \label{platino_gauged}%
\end{equation}
Contracting  Eq.~(\ref{platino_gauged}) by $\ \pi^{\alpha}$ and using
$\widehat{T}_{\alpha \beta}$ from Eq.~(\ref{T_ab}) results in
the first gauged constraint,
\begin{equation}
\pi\cdot\psi=\frac{e}{3m^{2}}\left[  -\gamma_{5}(\gamma^{\mu}\widetilde
{F}_{\beta\mu}\not \pi +\pi^{\mu}\widetilde{F}_{\mu\beta})+2iF_{\beta\mu}%
\pi^{\mu}\right]  \psi^{\beta}\,, \label{1st_gauged}%
\end{equation}
whereas contraction by $\gamma^{\alpha}$ leads to
\begin{equation}
\left(  1-\frac{e}{3m^{2}}\sigma^{\mu\nu}F_{\mu\nu}\right)  \gamma\cdot
\psi=\frac{e}{3m^{2}}\gamma^{\mu}\left(  \gamma_{5}\widetilde{F}_{\mu\beta
}+4iF_{\mu\beta}\right)  \psi^{\beta}. \label{2nd_gauged}%
\end{equation}
To obtain the latter equation we used the equivalent representation
for $T_{\alpha\beta}$
\begin{equation}
T_{\alpha\beta}=g_{\alpha\beta} \not \pi
-(\gamma_{\alpha}\pi_{\beta}+\gamma_{\beta}\pi_{\alpha}
-\gamma_{\alpha}\not \pi \gamma_{\beta})\, .
\label{T_ab_anders}
\end{equation}
In the $1/m$ expansion the gauged constraints read
\begin{align}
\pi\cdot\psi &  =0+\mathcal{O}(1/m^{2})\,,\nonumber\\
\gamma\cdot\psi &  =0+\mathcal{O}(1/m^{2}). \label{LO_GaCo}%
\end{align}
Substitution of Eqs.~(\ref{LO_GaCo}) into Eq.~(\ref{platino_gauged}) amounts
to our prime result-- the explicit form of the gauged
equation (\ref{gauged_hidden}):
\begin{equation}
(\pi^{2}-m^{2})\psi_{\alpha}+\frac{2i}{3}eF_{\alpha\beta}\psi^{\beta}%
+\frac{e}{6}F^{\mu\nu}\sigma_{\mu\nu}\psi_{\alpha
}+\frac{ie}{3}\gamma_{\alpha}\gamma^{\eta}F_{\eta\beta}\psi^{\beta}+\frac
{e}{3}\gamma_{5}\widetilde{F}_{\alpha\beta}\psi^{\beta}=0. 
\label{gauged_NDD...}%
\end{equation}
In recalling that the magnetic interaction is described by the covariant term
\begin{equation}
M_{\mu\nu}F^{\mu\nu}\,, \label{MI_cov}%
\end{equation}
where $M_{\mu\nu\text{ }}$ stand for the
Lorentz group generators (see Appendix), 
one finds that for Dirac theory the magnetic
interaction is given by
\begin{equation}
\mathcal{L}^{(s=1/2)}_{mag}=
\frac{ge}{2}\overline{\psi}_{a}(\frac{\sigma_{\mu\nu}}%
{2})_{ab}\psi_{b}F^{\mu\nu}\,. \label{Dirac_MgInt}%
\end{equation}
Similarly, for Proca particles one encounters%
\begin{equation}
\mathcal{L}^{(s=1)}_{mag}=
\frac{ge}{2}W_{\alpha}^{\dagger}(L^{\mu\nu})^{\alpha\beta
}W_{\beta}F^{\mu\nu}=igeW_{\mu}^{\dagger}W_{\nu}F^{\mu\nu}\,,
\label{Proca_MgInt}%
\end{equation}
where we used $(L^{\mu\nu})^{\alpha\beta}=i(g^{\mu\alpha}g^{\nu\beta}%
-g^{\mu\beta}g^{\nu\alpha})$. In a similar way we expect for any
representation
\begin{equation}
\mathcal{L}_{mag}=\frac{ge}{2}\overline{\psi}_{A}(M_{\mu\nu})_{AB}\psi
_{B}F^{\mu\nu}\,, \label{anyJ_MgInt}%
\end{equation}
where capital Latin indices stand for the complete set of quantum numbers
that label $M_{\mu\nu}$  in the internal space of $\psi$ and
its adjoint, respectively. In particular for the Rarita-Schwinger space under
consideration
\begin{equation}
\mathcal{L}^{(s=3/2)}_{mag}=
\frac{ge}{2}\overline{\psi}_{a}^{\alpha}(M_{\mu\nu}%
)_{\alpha\beta;ab}\psi_{b}^{\beta}F^{\mu\nu}. \label{3_2_MgInt}%
\end{equation}
As long as
\begin{equation}
(M_{\mu\nu})_{\alpha\beta;ab}=(L_{\mu\nu})_{\alpha\beta}\delta_{ab}%
+g_{\alpha\beta}(\frac{1}{2}\sigma^{\mu\nu})_{ab},
\end{equation}
and in suppressing Dirac indices leads to the following interacting
lagrangian
\begin{equation}
\mathcal{L}^{(s=3/2)}_{mag}=\frac{ge}{2}\overline{\psi}^{\alpha}[(L_{\mu\nu}%
)_{\alpha\beta}+g_{\alpha\beta}\frac{1}{2}\sigma^{\mu\nu}]\psi^{\beta}%
F^{\mu\nu}=ige\overline{\psi}_{\mu}\psi_{\nu}F^{\mu\nu}+\frac{ge}{2}%
\overline{\psi}^{\alpha}[g_{\alpha\beta}\frac{1}{2}\sigma^{\mu\nu}]\psi
^{\beta}F^{\mu\nu}\,. \label{Int_Lagr}%
\end{equation}
In comparing the latter equation to Eq.~(\ref{gauged_NDD...}) 
allows to conclude the
gyromagnetic factor as
\begin{equation}
g=\frac{2}{3}=\frac{1}{s}, \quad s=\frac{3}{2}\, . \label{Urrra}%
\end{equation}
In conclusion, Eq.~(\ref{platino_gauged}) predicts the correct value for the
gyromagnetic factor in accordance with Belinfante's conjecture
\cite{Belinfante} on the inverse proportionality between $g_{s}$ and spin.
Moreover, Eq.~(\ref{gauged_NDD...}) is manifestly hyperbolic and causal
as the second order time derivative enter the equation only via its 
Klein-Gordon building block. A further advantage of 
Eq.~(\ref{gauged_NDD...}) is that the electromagnetic 
interaction of the $s=3/2$ sector is already turned on at leading order in the 
$\frac{1}{m}$ expansion. 
Notice that to same order the Rarita-Schwinger equation (\ref{VZ_gauged})
gauged by Velo and Zwanziger reduces to the mere gauged Dirac equation
and leaves the coupling of the spin 3/2 degrees of freedom unspecified.
Within the Rarita-Schwinger framework the gyromagnetic factor is 
extracted at the non-relativistic level 
\cite{MC}, \cite{Weda},\cite{Nozawa} or from calculating  pion-nucleon
bremsstrahlung and a subsequent comparison to low energy theorems
\cite{Psct_EM}. Our equation (\ref{gauged_NDD...}) has the advantage to allow
identification of the magnetic coupling in a fully covariant fashion.

Incorporation of the auxiliary conditions beyond leading order is in work.
In Ref.~\cite{NK03} we studied the propagation of the wave fronts
of the complete ${\mathcal W}^{2}$ eigenvalue problem and 
delivered the proof that at least in the basis where
${\mathcal W}^{2}$ diagonalizes, the associated spin 3/2 equation of motion
is free from the Velo-Zwanziger problem. However, in this basis,
one no longer has separation between Lorentz and Dirac indices.
In case  the Velo-Zwanziger problem turns out to be related
to the particular representation choice in Eq.~(\ref{drct_prdct}) 
then we at least succeeded in pushing away the inconsistencies to order 
${\mathcal O} \left(1/m^{2}\right)$ 
in the $1/m$ expansion. 

\section{Summary.}
Our suggested solution to the problem of 
the covariant and consistent description of spin 3/2
within an electromagnetic environment is
the Klein-Gordon-like equation (\ref{platino}) 
(gauged (\ref{gauged_NDD...})) following
from the ${\mathcal W}^{2}$ eigenvalue problem in $\psi_{\mu}$ and
which is (i) fully covariant, (ii) parameter free, 
(iii) hyperbolic and causal, (iv) covariantly coupled to the
electromagnetic field by a gyromagnetic factor of
$g_{3/2}=2/3$.
Though we argued at the level of the vector-spinor, our approach
in being based upon the complete 
${\mathcal W}^{2}$ eigenvalue problem-- a quantity that 
is well defined in any
representation space-- obviously 
can be generalized to any totally symmetric 
rank-$(s-\frac{1}{2})$ Lorentz tensor 
with Dirac spinor components, $\psi_{ \mu_{1}...\mu_{s-\frac{1}{2}} }$, 
and thereby to any fractional spin $s$.
Compared to the Dirac-like spin 3/2  equations, the advantage of
the Klein-Gordon like one is to have
achieved  a parameter free description and a 
covariant coupling to the electromagnetic field already 
to leading order in the  $1/m$ expansion.
We expect the proposed scheme to help improving 
description of various processes containing spin 3/2 fermions.

\section*{Acknowledgments}

Work supported by Consejo Nacional de Ciencia y Tecnologia (CONACyT) Mexico
under projects 37234-E and C01-39820. We extend our special thanks to
Selim Gomez-Avila for joining research on ${\mathcal W}^{2}$ induced
high spin description and sharing
unselfishly the experience he won while exploring spin 2.

\section{Appendix 1}

In this appendix we collect conventions and some results on the symmetry of
spacetime under rotations, boosts and translations- transformations that
constitute the Poincar\'{e} group for which the squared Pauli-Lubanski vector
is a Casimir invariant. In terms of the Poincar\'{e} group generators,
$M_{\mu\nu}$ and $P_{\eta}$ and their algebra \cite{WKT}
\begin{align}
\left[  M_{\mu\nu},M_{\alpha\beta}\right]   &  =-i(g_{\mu\alpha}M_{\nu\beta
}-g_{\mu\beta}M_{\nu\alpha}+g_{\nu\beta}M_{\mu\alpha}-g_{\nu\alpha}M_{\mu
\beta})\,,\label{M_comm}\\
\left[  M_{\alpha\beta},p_{\mu}\right]   &  =-i(g_{\mu\alpha}p_{\beta}%
-g_{\mu\beta}p_{\alpha}),\quad\left[  p_{\mu},p_{\nu}\right]
=0\,,\label{M_P_comm}%
\end{align}
where $g_{\mu\nu}$=diag$(1,-1,-1,-1)$ is the metric tensor, the
Pauli--Lubanski (PL) vector is defined as
\begin{equation}
{\mathcal W}_{\mu}=\frac{1}{2}\epsilon_{\mu\nu\alpha\beta}M^{\nu\alpha}p^{\beta
}\,,\label{paulu}%
\end{equation}
with $\epsilon_{0123}=1$. This operator can be shown to satisfy the following
commutation relations
\begin{equation}
\lbrack M_{\mu\nu},{\mathcal W}_{\alpha}]=
-i(g_{\alpha\mu}{\mathcal W}_{\nu}-g_{\alpha\nu}
{\mathcal W}_{\mu
}),\qquad\lbrack {\mathcal W}_{\alpha},p_{\mu}]=0,
\qquad\lbrack {\mathcal W}_{\alpha},{\mathcal W}_{\beta
}]=-i\epsilon_{\alpha\beta\mu\nu}{\mathcal W}^{\mu}p^{\nu}\,,\label{conmrelpl}%
\end{equation}
\noindent i.e. it transforms as a four-vector under Lorentz transformations.
Moreover, its square commutes with all the generators and is a group
invariant. For this reason elementary particles are required to transform
invariantly under the action of ${\mathcal W}^{2}$ and to be 
labeled by the ${\mathcal W}^{2}$
eigenvalues next to those of $p^{2}$. The representation spaces of interest
for the present work are the four-vector and the Dirac-spinor. Their
respective Pauli-Lubanski vectors, $W_{\mu}$, and $w_{\mu}$ are found as
\begin{align}
\left[  W_{\mu}\right]  _{\alpha}^{\quad\beta} &  =i\epsilon_{\mu\alpha
\quad\sigma}^{\quad\beta}p^{\sigma}\,,\label{PL_hh}\\
w_{\mu} &  =\frac{i}{2}\gamma_{5}\sigma_{\mu\rho}p^{\rho}\,.\label{PL_Dir}%
\end{align}
To obtain above expressions we substituted for the generators in the
four-vector and the Dirac-spinor in Eq.~(\ref{paulu}) $\left[  M^{\mu\nu
}\right]  _{\alpha}\,^{\beta}=i\left(  g_{\alpha}\,^{\mu}g^{\nu\beta}%
-g^{\mu\beta}g_{\alpha}\,^{\nu}\right)  $, and $M_{\mu\nu}=\frac{1}{2}%
\sigma_{\mu\nu}$, respectively. The Casimir invariants in the four-vector and
Dirac-spinor spaces are now calculated as
\begin{equation}
\left[  W^{2}\right]  _{\alpha}^{\quad\beta}=-2\left(  g_{\alpha}^{\beta}%
p^{2}-p^{\beta}p_{\alpha}\right)  \,,\quad w^{2}=-\frac{3}{4}p^{2}%
\,.\label{W2_hh}%
\end{equation}
In particular Proca's equation is just the corresponding 
eigenvalue equation for the spin $1$ subspace%
\begin{equation}
\left[  W^{2}\right]  _{\alpha}^{\quad\beta}A_{\beta}=-2m^{2}A_{\alpha}\, .
\end{equation}

\end{document}